\documentclass[12pt,draftclsnofoot, perreview, onecolumn]{IEEEtran}

\usepackage{amsfonts}
\usepackage{amsmath}
\usepackage{mathrsfs}
\usepackage{multirow}
\usepackage{booktabs}
\usepackage{setspace}
\usepackage{amssymb}
\usepackage{mdwmath}
\usepackage{balance}
\usepackage{color}
\usepackage{url}
\usepackage{citesort}
\usepackage{threeparttable}
\usepackage{ifpdf}
\usepackage[dvips]{graphicx}
\graphicspath{{../}}
\DeclareGraphicsExtensions{.pdf,.jpeg,.png,.eps}
\ifCLASSOPTIONcompsoc
  \usepackage[tight,normalsize,sf,SF]{subfigure}
\else
  \usepackage[tight,footnotesize]{subfigure}
\fi

\newcommand{\mathbfit}[1]{\mbox{\boldmath$#1$\unboldmath}}
\newcommand{\diag}{\mbox{diag}}

\newtheorem{algorithm}{\textbf{Algorithm}}

\newtheorem{example}{\textbf{Example}}

\ifCLASSOPTIONonecolumn
\newcommand{\figwidth}{0.75\textwidth}
\fi

\ifCLASSOPTIONtwocolumn
\newcommand{\figwidth}{0.48\textwidth}
\fi

\begin{document}

\title{Block Markov Superposition Transmission: Construction of Big Convolutional Codes from Short Codes}

\author{Xiao~Ma,~\IEEEmembership{Member,~IEEE,}
        Chulong~Liang,
        Kechao~Huang,
        and~Qiutao~Zhuang
\thanks{This work is supported by the 973 Program (No.2012CB316100) and the NSF~(No.61172082) of China. This paper was presented in part at the IEEE International Symposium on Information Theory, 2013.}
\thanks{The authors are with the Department of Electronics and Communication
Engineering, Sun Yat-sen University, Guangzhou 510006, China (e-mail: maxiao@mail.sysu.edu.cn, lchul@mail2.sysu.edu.cn,
hkech@mail2.sysu.edu.cn, zhuangqt@mail2.sysu.edu.cn).}}


\maketitle

\begin{abstract}
A construction of big convolutional codes from short codes called block Markov superposition transmission~(BMST) is proposed. The BMST is very similar to superposition block Markov encoding~(SBME), which has been widely used to prove multiuser coding theorems. The encoding process of BMST can be as fast as that of the involved short code, while the decoding process can be implemented as an iterative sliding-window decoding algorithm with a tunable delay. More importantly, the performance of BMST can be simply lower-bounded in terms of the transmission memory given that the performance of the short code is available. Numerical results show that, 1)~the lower bounds can be matched with a moderate decoding delay in the low bit-error-rate~(BER) region, implying that the iterative sliding-window decoding algorithm is near optimal; 2)~BMST with repetition codes and single parity-check codes can approach the Shannon limit within 0.5 dB at BER of $10^{-5}$ for a wide range of code rates; and 3)~BMST can also be applied to nonlinear codes.
\end{abstract}


\section{Introduction}\label{sec:Introduction}
Convolutional codes, first introduced by Elias~\cite{Elias55}, have been used in various communication systems~\cite{Costello98}, such as space communication, data transmission, digital audio/video transmission, and mobile communication. In these systems, only convolutional codes with short constraint lengths are implemented due to the fact that the decoding complexity of the Viterbi algorithm~\cite{Viterbi67} grows exponentially with the constraint length.\footnote{On the Galileo mission to Jupiter, a convolutional code was implemented with the big Viterbi decoder (BVD) over a trellis of $2^{14} = 16384$ states~\cite{Costello98}~\cite{Collins92}.} Constructing~(decodable) convolutional codes with long constraint length~(referred to as {\em big} convolutional codes in this paper) is of interest both in theory and in practice.

It is an old subject to construct long codes from short codes~\cite{Macwilliams77}. Here, by {\em short codes}, we mean block codes with short code lengths or convolutional codes with short constraint lengths.  Product codes~\cite{Elias54}, presented by Elias in 1954, may be the earliest method for constructing long codes with short codes. An $[n_1n_2, k_1k_2]$ product code is formed by an $[n_1, k_1]$ linear code~$\mathscr{C}_1$ and an $[n_2, k_2]$ linear code~$\mathscr{C}_2$. Each codeword of the product code is a rectangular array of $n_1$ columns and $n_2$ rows in which each row is a codeword in $\mathscr{C}_1$ and each column is a codeword in $\mathscr{C}_2$. In 1966, Forney proposed a class of codes, called concatenated codes~\cite{Forney66}. Typically, a concatenated code investigated by Forney consists of a relatively short code as an inner code and a relatively long algebraic code as an outer code. In 1993, Berrou~{\em et al} invented turbo codes~\cite{Berrou93}, by which researchers have been motivated to construct capacity-approaching codes.  The original turbo code~\cite{Berrou93} consists of two convolutional codes which are parallelly concatenated by a pseudo-random interleaver, and hence is also known as a parallel concatenated convolutional code~(PCCC)~\cite{Benedetto96}. Since the invention of turbo codes, concatenations of simple interleaved codes have been proved to be a powerful approach to design iteratively decodable capacity-approaching codes~\cite{Benedetto98,Divsalar98,Ping01,Abbasfar07,Tong12}. Another class of capacity-approaching codes, namely, low-density parity-check~(LDPC) codes, which were proposed in the early 1960s and rediscovered after the invention of turbo codes, can also be considered~(from the aspect of decoding) as concatenations of interleaved single parity-check codes and repetition codes~\cite{Gallager63,Mackay99,Richardson01b,Kou01,Lan07,Felstrom99,Pusane11}.

In this paper, we present more details on the recently proposed block Markov superposition transmission~(BMST)~\cite{Ma13}, which is a construction of big convolutional codes from short codes. The BMST is very similar to superposition block Markov encoding~(SBME), which has been widely used to prove multiuser coding theorems. The method of SBME was first introduced for the multiple-access channel with feedback by Cover and Leung~\cite{Cover81} and successfully applied by Cover and El Gamal~\cite{Cover79} for the relay channel. The idea behind SBME in the single-relay system can be briefly summarized as follows~\cite{Willems85}.

Assume that the data are equally grouped into $B$ blocks. Initially, the source broadcasts a codeword that corresponds to the first data block. Since the code rate is higher than the capacity of the link from the source to the destination, the destination is not able to recover the data reliably. Then the source and the relay cooperatively transmit more information about the first data block. In the meanwhile, the source ``superimposes" a codeword that corresponds to the second data block. Finally, the destination is able to reliably recover the first data block from the two successive received blocks. After removing the effect of the first data block, the system returns to the initial state. This process iterates $B+1$ times until all $B$ blocks of data are sent successfully.

We apply a similar strategy to the point-to-point communication system. We assume that the transmitter uses a short code. Initially, the transmitter sends a codeword that corresponds to the first data block. Since the short code is {\em weak}, the receiver is unable to recover  reliably the data from the current received block. Hence the transmitter transmits the codeword~(in its interleaved version) one more time. In the meanwhile, a fresh codeword that corresponds to the second data block is superimposed on the second block of transmission. Finally, the receiver recovers the first data block from the two successive received blocks. After removing the effect of the first data block, the system returns to the initial state. This process iterates $B+1$ times until all $B$ blocks of data are sent successfully. In practice, the receiver may use an iterative sliding-window decoding algorithm. The system performance can be analyzed in terms of the transmission memory and the input-output weight enumerating function~(IOWEF) of the BMST system, which can be computed from that of the short code using a trellis-based algorithm. Simulation results verify our analysis and show that remarkable coding gain can be obtained.

The rest of this paper is organized as follows. We present the encoding algorithm of the BMST system and derive its generator matrix and parity-check matrix in Section~\ref{sec:Encoding}. In Section~\ref{sec:Deocding}, we focus on the decoding algorithms of the BMST system. In Section~\ref{sec:PerformanceAnalysis}, the performance of the BMST system is analyzed with a simple lower bound by assuming a genie-aided decoder and an upper bound with the help of the IOWEF. Numerical results are presented in Section~\ref{sec:Result}. Section~\ref{sec:Universality} discusses the universality of the BMST. Section~\ref{sec:Conclusion} concludes this paper.

\section{Block Markov Superposition Transmission}\label{sec:Encoding}
\subsection{Encoding Algorithm}\label{subsec:Encoding}
We focus on binary codes in this paper. For a rate $R = k/n$ binary convolutional code, information sequence $\mathbfit{u} = \left( \mathbfit{u}^{(0)}, \mathbfit{u}^{(1)}, \cdots \right) = \left(  u_0^{(0)}, \cdots, u_{k-1}^{(0)}, u_0^{(1)}, \cdots, u_{k-1}^{(1)}, \cdots \right)$ is encoded into code sequence $\mathbfit{c} = \left( \mathbfit{c}^{(0)}, \mathbfit{c}^{(1)}, \cdots \right) = \left(  c_0^{(0)}, \cdots, c_{n-1}^{(0)}, c_0^{(1)}, \cdots, c_{n-1}^{(1)}, \cdots \right)$. The encoding process is initialized by setting $\mathbfit{u}^{(t)} = 0$ for $t<0$ and computes for $t \geq 0$ as shown in~\cite{Johannesson99}
\begin{equation}\label{eq:generalConvCodeMatrixForm}
\mathbfit{c}^{(t)} = \mathbfit{u}^{(t)}\mathbfit{G}_0 + \mathbfit{u}^{(t-1)}\mathbfit{G}_1 + \cdots + \mathbfit{u}^{(t-m)}\mathbfit{G}_m,
\end{equation}
where $\mathbfit{G}_i$ $(0 \leq i \leq m)$ is a binary $k \times n$ matrix and $m$ is called the encoder \emph{memory}.

\begin{figure}[t]
   \centering
   \includegraphics[width=\figwidth]{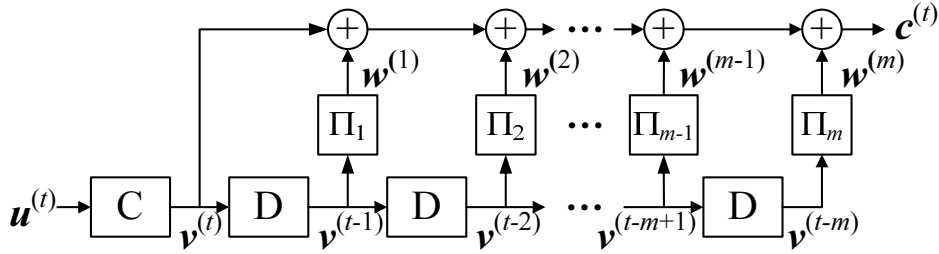}
   \caption{Encoding structure of a BMST system with memory $m$.}
   \label{fig:encoder}\vspace{-0.5cm}
\end{figure}
In this paper, we propose a special class of convolutional codes by setting $\mathbfit{G}_0 = \mathbfit{G}$ and $\mathbfit{G}_i = \mathbfit{G}\mathbfit{\varPi}_i$, where $\mathbfit{G}$ is the generator matrix of a binary linear code $\mathscr{C}[n, k]$ of dimension $k$ and length $n$ and  $\mathbfit{\varPi}_i$~$(1 \leq i \leq m)$ is a permutation matrix of size $n\times n$. The code $\mathscr{C}[n, k]$ is referred to as the \emph{basic code} in this paper for convenience. Let $\mathbfit{u}^{(0)}$, $\mathbfit{u}^{(1)}$, $\cdots$, $\mathbfit{u}^{(L-1)}$ be $L$ blocks of data to be transmitted, where $\mathbfit{u}^{(t)} \in \mathbb{F}_2^k$. The encoding algorithm with memory $m$ is described as follows, see Fig.~\ref{fig:encoder} for reference, where the permutation matrix $\mathbfit{\varPi}_i$ is implemented as its corresponding interleaver of size $n$.
\begin{algorithm}{Encoding of BMST}\label{alg:encoding}
\begin{itemize}
  \item{\bf{Initialization}:} \label{step:encoding_initialize} For $t < 0$, set $\mathbfit{v}^{(t)} = \mathbfit{0} \in \mathbb{F}_2^n$.
  \item{\bf{Loop}:} \label{step:encoding_iteration} For $t = 0$, $1$, $\cdots$, $L-1$,
        \begin{enumerate}
          \item Encode $\mathbfit{u}^{(t)}$ into $\mathbfit{v}^{(t)} \in \mathbb{F}_2^n$ by the encoding algorithm of the basic code $\mathscr{C}$;
          \item For $1\leq i \leq m$, interleave $\mathbfit{v}^{(t-i)}$ by the $i$-th interleaver $\mathbfit{\varPi}_{i}$ into $\mathbfit{w}^{(i)}$;
          \item Compute $\mathbfit{c}^{(t)} = \mathbfit{v}^{(t)} + \sum_{1\leq i \leq m} \mathbfit{w}^{(i)}$, which is taken as the $t$-th block of transmission.
        \end{enumerate}
  \item{\bf{Termination}:} \label{step:encoding_termination}
        For $t = L$, $L+1$, $\cdots$, $L+m-1$, set $ \mathbfit{u}^{(t)} = \mathbfit{0} \in \mathbb{F}_2^k$ and compute $\mathbfit{c}^{(t)}$ following Step.~{\bf Loop}.
  \end{itemize}
\end{algorithm}

\textbf{Remarks.} The code rate is $\frac{kL}{n(L+m)}$, which is slightly less than that of the basic code $\mathscr{C}$. However, the rate loss is negligible for large $L$. Also notice that interleaving $\mathbfit{v}^{(t-i)}$ into $\mathbfit{w}^{(i)}$ and encoding $\mathbfit{u}^{(t)}$ into $\mathbfit{v}^{(t)}$ can be implemented in parallel. Therefore, the encoding process for the BMST system can be almost as fast as the encoding process for the basic code $\mathscr{C}$ given that sufficient hardware resources are available.

\subsection{Algebraic Description of BMST}\label{subsec:AlgebraStructure}
Unlike commonly accepted classical convolutional codes, the codes specified by the BMST system typically have large $k$ and (hence) large constraint lengths. From a practical point of view, we are mainly concerned with the terminated BMST. In this case, the BMST system can be treated as a linear block code $\mathscr{C}[n(L+m), kL]$. In the following, we present for integrity the generator matrix and the parity-check matrix of the BMST system although we have not found their usefulness in describing both the encoding algorithm and the decoding algorithm.

Let $\mathbfit{G}$ and $\mathbfit{H}$ be the generator matrix and the parity-check matrix of the basic code, respectively. Let $\mathbfit{\varPi}_i$, $1 \leq i \leq m$, be the $m$ involved permutation matrices. The generator matrix of the BMST system is given by
\begin{equation}\label{eq:GBMST}
\mathbfit{G}_{{\small \rm BMST}} =
  \diag\{\underbrace{\mathbfit{G}, \cdots, \mathbfit{G}}_L \} \mathbfit{\varPi},
\end{equation}
where $\diag\left\lbrace \mathbfit{G}, \cdots, \mathbfit{G} \right\rbrace$ is a block diagonal matrix with $\mathbfit{G}$ on the diagonal and $\mathbfit{\varPi}$ is a block upper banded matrix~(consisting of $L$ rows and $L+m$ columns of sub-blocks) as shown below,
\begin{equation}\label{eq:G_BMST}
\mathbfit{\varPi} =
  \left[
        \begin{array}{ccccccccccccc}
           \mathbfit{I}       & \mathbfit{\varPi}_{1} & \cdots &
           \mathbfit{\varPi}_{m} &                    &  \\
                              & \mathbfit{I}       & \mathbfit{\varPi}_{1} &
           \cdots             & \mathbfit{\varPi}_{m} &  \\
                              &                     & \ddots &
           \ddots             & \ddots              & \ddots  \\
                              &                     &        &
           \mathbfit{I}       & \mathbfit{\varPi}_{1}  & \cdots &
           \mathbfit{\varPi}_{m} & \\
                              &                     &        &
                              & \mathbfit{I}        & \mathbfit{\varPi}_{1} &
           \cdots             & \mathbfit{\varPi}_{m} \\
        \end{array}
  \right].
\end{equation}
Apparently, $\mbox{Rank}(\mathbfit{G}_{{\small \rm BMST}}) = kL$ since $\mbox{Rank}(\mathbfit{G}) = k$ and $\mathbfit{\varPi}$ is of full rank. From the generator matrix $\mathbfit{G}_{{\small \rm BMST}}$, we can see that the minimum Hamming weight of the BMST system is at least as twice as that of the basic code.

To derive the parity-check matrix of the BMST system, we define recursively a sequence of matrices as $\mathbfit{P}_0 = \mathbfit{I}$~(the identity matrix of order $n$) and $\mathbfit{P}_{t} = \sum_{1\leq \ell \leq m} \mathbfit{P}_{t-\ell} \mathbfit{\varPi}_{\ell}$ for $t \geq 1$, where $\mathbfit{P}_t$ for $t < 0$ are initialized to be the zero matrix of order $n$. From the fact that $\mathbfit{c}^{(t)} = \mathbfit{v}^{(t)} + \sum_{\ell=1}^{m} \mathbfit{v}^{(t-\ell)} \mathbfit{\varPi}_\ell$, we conclude that $\mathbfit{v}^{(t)}$ can be found recursively from $\mathbfit{c}^{(t)}$ as  $\mathbfit{v}^{(t)} = \mathbfit{c}^{(t)} + \sum_{\ell=1}^{m} \mathbfit{v}^{(t-\ell)} \mathbfit{\varPi}_\ell$. Equivalently, we have $(\mathbfit{v}^{(0)}, \mathbfit{v}^{(1)}, \cdots, \mathbfit{v}^{(L+m-1)}) = (\mathbfit{c}^{(0)}, \mathbfit{c}^{(1)}, \cdots, \mathbfit{c}^{(L+m-1)}) \mathbfit{P}$, where $\mathbfit{P}$ is block upper triangular matrix~(consisting of $L+m$ rows and $L+m$ columns of sub-blocks) as shown below,
\begin{equation}\label{eq:p_matrix_simple}
  \mathbfit{P} = \left [
        \begin{array}{cccccccc}
           \mathbfit{I} & \mathbfit{P}_{1} & \mathbfit{P}_{2} & \cdots & \mathbfit{P}_{L+m-1} \\
           & \mathbfit{I} & \mathbfit{P}_{1} & \cdots & \mathbfit{P}_{L+m-2} \\
           & & \ddots & \ddots & \vdots \\
           & & & \mathbfit{I} & \mathbfit{P}_{1} \\
           & &  &  & \mathbfit{I} \\
        \end{array}
   \right ].
\end{equation}
Since $\mathbfit{v}^{(t)}$ is a codeword in the basic code and $\mathbfit{v}^{(t)} = \mathbfit{0}$ for $t\geq L$, we know
\begin{equation}
\left(\mathbfit{c}^{(0)}, \mathbfit{c}^{(1)}, \cdots, \mathbfit{c}^{(L+m-1)}\right)\mathbfit{P} \cdot \diag\{\underbrace{\mathbfit{H}^{\rm T}, \cdots, \mathbfit{H}^{\rm T}}_L, \underbrace{\mathbfit{I}, \cdots, \mathbfit{I}}_m \} = \mathbf{0},
\end{equation}
where the superscript T denotes ``transpose". Now we claim that the parity-check matrix of the BMST system is given by
\begin{equation} \label{eq:HBMST}
\mathbfit{H}_{\mbox{\small BMST}} = \diag\{\underbrace{\mathbfit{H}, \cdots, \mathbfit{H}}_L, \underbrace{\mathbfit{I}, \cdots, \mathbfit{I}}_m\} \mathbfit{P}^{\rm T}.
\end{equation}
This is justified by noting that $\mbox{Rank}\left(\mathbfit{H}_{\mbox{\small BMST}}\right) = (n-k)L + nm$.

\section{Iterative Sliding-window Decoding Algorithm}\label{sec:Deocding}
\subsection{Notation of Normal Graphs}\label{subsec:NormalGraphNotation}
\begin{figure}[t]
  \centering
  \includegraphics[width=\figwidth]{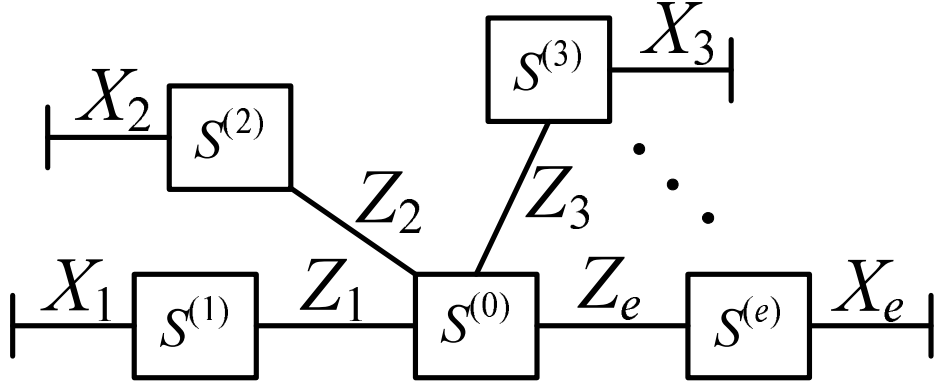}
  \caption{A normal graph of a general (sub)system.}
  \label{fig:GeneralNormalGraph}
\end{figure}
Before describing the decoding algorithm, we introduce the message processing/passing algorithm over a general normal graph~\cite{Forney01}. The notation is closely related to that used in~\cite{Ma04,Ma12}. As shown in Fig.~\ref{fig:GeneralNormalGraph}, a general normal graph can be used to represent a system, where vertices represent subsystems and edges represent variables. All edges~(variables) connecting to a vertex~(subsystem) must satisfy the specific constraints of the subsystem. For example, the subsystem $S^{(0)}$ is connected to $S^{(j)}$ via $Z_j$, and the subsystem $S^{(j)}$ is potentially connected to other system via a half edge $X_j$. Associated with each edge is a {\em message} that is defined in this paper as the probability mass function~(pmf) of the corresponding variable. We focus on random variables defined over $\mathbb{F}_2$. We use the notation $P_{Z_{j}}^{(S^{(0)} \rightarrow S^{(j)})}(z), z \in \mathbb{F}_2$ to denote the message from vertex $S^{(0)}$ to vertex $S^{(j)}$. Suppose that all messages $P_{Z_{j}}^{(S^{(j)} \rightarrow S^{(0)})}(z), z \in \mathbb{F}_2$ are available. Then, the vertex $S^{(0)}$, as a {\em message processor}, delivers the outgoing message with respect to any given $Z_{j}$ by computing the likelihood function
\begin{equation}\label{eq:message_processing}
    P_{Z_{j}}^{(S^{(0)} \rightarrow S^{(j)})}(z) \propto
    \mbox{Pr} \left \lbrace S^{(0)}
    \mbox{is satisfied} | Z_{j} = z \right \rbrace,
    z \in \mathbb{F}_2.
\end{equation}
Because the computation of the likelihood function is irrelevant to the incoming message \\ $P_{Z_{j}}^{(S^{(j)} \rightarrow S^{(0)})}(z)$, we claim that $P_{Z_{j}}^{(S^{(0)} \rightarrow S^{(j)})}(z)$ is exactly the so-called {\em extrinsic message}. For simplicity, if two subsystems share multiple variables of the same type, the corresponding edges can be merged into one edge. Such an edge represents a sequence of random variables, whose messages are then collectively written in a sequence. Notice that such a simplified representation is just for the convenience of describing the message passing. For message processing, any edge that represents multiple random variables must be treated as multiple separated edges.

\begin{figure}[t]
  \centering
  \includegraphics[width=\figwidth]{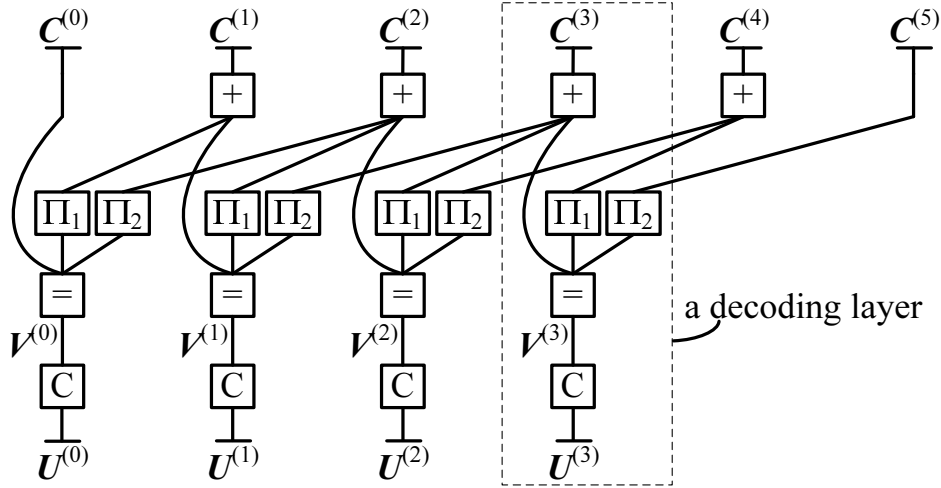}
  \caption{The normal graph of a BMST system with $L = 4$ and $m = 2$.}
  \label{fig:decoder}
\end{figure}
Fig.~\ref{fig:decoder} shows the normal graph of a BMST system with $L=4$ and $m=2$. There are four types of nodes in the normal graph of the BMST system, and each edge represents a sequence of random variables.
\begin{itemize}
  \item{\emph{Node} \fbox{C}:} The node \fbox{C} represents the constraint that $\mathbfit{V}^{(t)}$ must be a codeword of $\mathscr{C}$ that corresponds to $\mathbfit{U}^{(t)}$. In practice, $\mathbfit{U}^{(t)}$ is usually assumed to be independent and uniformly distributed over $\mathbb{F}_2^k$. Assume that the messages associated with $\mathbfit{V}^{(t)}$ are available from the node \fbox{=}. The node \fbox{C} performs a soft-in-soft-out~(SISO) decoding algorithm to compute the extrinsic messages. The extrinsic messages associated with $\mathbfit{V}^{(t)}$ are fed back to the node \fbox{=}, while the extrinsic messages associated with $\mathbfit{U}^{(t)}$ can be used to make decisions on the transmitted data.

  \item{\emph{Node} \fbox{=}:} The node \fbox{=} represents the constraint that all connecting variables must take the same realizations. The message processing/passing algorithm of the node \fbox{=} is the same as that of the variable node in an LDPC code.

  \item{\emph{Node} \fbox{$\Pi_{i}$}:} The node \fbox{$\Pi_{i}$} represents the $i$-th interleaver, which interleaves or de-interleaves the input messages.

  \item{\emph{Node} \fbox{+}:} The node \fbox{+} represents the constraint that all connecting variables must be added up to zero over $\mathbb{F}_2$. The message processing/passing algorithm at the node \fbox{+} is similar to that at the check node in an LDPC code. The only difference is that the messages associated with the half edge are computed from the channel observations.
\end{itemize}

The normal graph of a BMST system can be divided into \emph{layers}, where each layer typically consists of a node of type \fbox{C}, a node of type \fbox{=}, $m$ nodes of type \fbox{$\Pi$}, and a node of type \fbox{+}, see Fig.~\ref{fig:decoder} for reference.

\subsection{Decoding Algorithm}\label{subsec:Deocding}
For simplicity, we assume that $\mathbfit{c}^{(t)}$ is modulated using binary phase-shift keying~(BPSK) with 0 and 1 mapped to $+1$ and $-1$, respectively and transmitted over an additive white Gaussian noise~(AWGN) channel, resulting in a received vector $\mathbfit{y}^{(t)}$. In more general settings, we assume that the {\it a~posteriori} probabilities ${\rm Pr}\{c_j^{(t)} = 0, 1 | \mathbfit{y}^{(t)}\}$ are computable\footnote{The computation in this step is irrelevant to the code constraints but depends only on the modulation and the channel.}, where $c_j^{(t)}$ is the $j$-th component of $\mathbfit{c}^{(t)}$.

After all $\mathbfit{y}^{(t)}$ for $0 \leq t \leq L+m-1$ are received, an \emph{iterative forward-backward decoding} can be implemented to obtain the decoding result $\mathbfit{\hat{u}}^{(t)}(0 \leq t \leq L-1)$. The algorithm is scheduled as follows.\vspace{0.1cm}
\begin{algorithm}{Iterative Forward-Backward Decoding of BMST}\label{alg:TotallyDecoding}
\begin{itemize}
  \item {\bf{Initialization}:} Considering only the channel constraint, compute the {\em a~posteriori} probabilities $P_{\mathbfit{C}^{(t)}}^{\left( | \rightarrow + \right)}\left( \mathbfit{c}^{(t)} \right)$ from the received vector $\mathbfit{y}^{(t)}$ for $0 \leq t \leq L+m-1$. All messages over the intermediate edges are initialized as uniformly distributed variables. Notice that $\mathbfit{u}^{(t)} = 0$ for $t < 0$ and $t \geq L$. Set a maximum iteration number $I_{\max} > 0$.

  \item {\bf{Iteration}:}\label{step:Iteration}
          For $I = 1$, $2$, $\cdots$, $I_{\max}$,
          \begin{enumerate}
            \item{\bf{\textit{Forward recursion}}:}\label{step:up2down_message_update}
                For $t = 0$, $1$, $\cdots$, $L+m-1$, the $t$-th layer performs a message processing/passing algorithm scheduled as
                \begin{equation*}
                    \fbox{+} \rightarrow \fbox{$\Pi$} \rightarrow \fbox{=} \rightarrow                        \fbox{C} \rightarrow \fbox{=} \rightarrow \fbox{$\Pi$} \rightarrow \fbox{+}.
                \end{equation*}
                In the above procedure, the message processor at each node takes as input all available messages from connecting edges and delivers as output extrinsic messages to connecting edges. Hence the messages from adjacent layers are utilized, and the messages to adjacent layers are updated by considering both the constraints in the $t$-th layer and the received vector $\mathbfit{y}^{(t)}$.

            \item{\bf{\textit{Backward recursion}}:}\label{step:down2up_message_update}
                    For $t=L+m-1$, $\cdots$, $1$, $0$, the $t$-th layer performs a message processing/passing algorithm scheduled as
                    \begin{equation*}
                       \fbox{+} \rightarrow \fbox{$\Pi$} \rightarrow \fbox{=} \rightarrow
                       \fbox{C} \rightarrow \fbox{=} \rightarrow \fbox{$\Pi$} \rightarrow \fbox{+}.
                    \end{equation*}

            \item{\bf{\textit{Hard decision}}:} \label{step:check_decoding_success}
              For $0 \leq t \leq L-1$, make hard decisions on $\mathbfit{u}^{(t)}$ resulting in $\mathbfit{\hat{u}}^{(t)}$.
              If certain conditions are satisfied, output $\mathbfit{\hat{u}}^{(t)}$ for $0 \leq t \leq L-1$ and exit the iteration. Stopping criteria are discussed in Section~\ref{subsec:StoppingCriteria}.
          \end{enumerate}
\end{itemize}
\end{algorithm}

The above algorithm~(Algorithm~\ref{alg:TotallyDecoding}) suffers from a large decoding delay for large $L$. Similar to the Viterbi algorithm in practical systems, we present the following {\em iterative sliding-window decoding} with a fixed decoding delay $d \geq 0$. In contrast to Algorithm~\ref{alg:TotallyDecoding}, the iterative sliding-window algorithm with decoding delay $d$ works over a subgraph consisting of $d+1$ consecutive layers, which delivers, at time $t+d$, as output the estimated data block $\mathbfit{\hat{u}}^{(t)}$ after $\mathbfit{y}^{(t+d)}$ is received and slides into the decoder. Usually, we take the decoding delay $d \geq m$. The schedule is described as follows.\vspace{0.1cm}
\begin{algorithm}{Iterative Sliding-window Decoding of BMST}\label{alg:decoding}
\begin{itemize}
  \item {\bf{Global initialization}:} Assume that $\mathbfit{y}^{(t)}, 0\leq t \leq d-1$ have been received. Considering only the channel constraint, compute the {\em a~posteriori} probabilities $P_{\mathbfit{C}^{(t)}}^{\left( | \rightarrow + \right)}\left( \mathbfit{c}^{(t)} \right)$ from the received vector $\mathbfit{y}^{(t)}$ for $0 \leq t \leq d-1$. All messages over the other edges within and connecting to the $t$-th layer~($0 \leq t \leq d-1$) are initialized as uniformly distributed variables. Set a maximum iteration number $I_{\max} > 0$.

  \item {\bf{Sliding-window decoding}:} For $t = 0$, $1$, $\cdots$, $L-1$,
      \begin{enumerate}
        \item{\bf{\textit{Local initialization}}:} \label{step:LocalInit}
           If $t+d \leq L+m-1$, compute the {\em a~posteriori} probabilities $P_{\mathbfit{C}^{(t+d)}}^{\left( | \rightarrow + \right)}\left( \mathbfit{c}^{(t+d)} \right)$ from the received vector $\mathbfit{y}^{(t+d)}$ and all messages over other edges within and
           connecting to the $(t+d)$-th layer are initialized as uniformly distributed variables.

        \item{\bf{\textit{Iteration}}:} \label{step:Iteration@SlidingWindow}
          For $I = 1$, $2$, $\cdots$, $I_{\max}$,
          \begin{enumerate}
            \item{{\em Forward recursion}:} \label{step:up2down_message_update}
                    For $i = 0$, $1$, $\cdots$, $\min\left(d, L+m-1-t\right)$, the $(t+i)$-th layer performs a message processing/passing algorithm scheduled as
                    \begin{equation*}
                        \fbox{+} \rightarrow \fbox{$\Pi$} \rightarrow \fbox{=} \rightarrow
                        \fbox{C} \rightarrow \fbox{=} \rightarrow \fbox{$\Pi$} \rightarrow \fbox{+}.
                    \end{equation*}

            \item{{\em Backward recursion}:} \label{step:down2up_message_update}
                    For $i=\min\left(d, L+m-1-t\right)$, $\cdots$, $1$, $0$, the $(t+i)$-th layer performs a message processing/passing algorithm scheduled as
                    \begin{equation*}
                       \fbox{+} \rightarrow \fbox{$\Pi$} \rightarrow \fbox{=} \rightarrow
                       \fbox{C} \rightarrow \fbox{=} \rightarrow \fbox{$\Pi$} \rightarrow \fbox{+}.
                    \end{equation*}

            \item{\bf{\textit{Hard decision}}:} \label{step:check_decoding_success}
              Make hard decisions on $\mathbfit{u}^{(t)}$ resulting in $\mathbfit{\hat{u}}^{(t)}$. If certain conditions are satisfied, output $\mathbfit{\hat{u}}^{(t)}$ and exit the iteration. Stopping criteria are discussed in Section~\ref{subsec:StoppingCriteria}.
        \end{enumerate}
        \item{\bf{\textit{Cancelation}}:} \label{step:Cancelation}
         Remove the effect of $\mathbfit{\hat{v}}^{(t)}$ on all layers by updating the \emph{a~posteriori}~probabilities as
            \begin{equation}\label{eq:cacelation}
                P_{C^{(t+i)}_j}^{(| \rightarrow +)}\left( a \right) \leftarrow \sum\limits_{b \in \mathbb{F}_2} P_{C^{(t+i)}_j}^{(| \rightarrow +)}\left( b \right) P_{W_j^{(i)}}^{(\Pi_{i}\rightarrow +)}\left( a+b \right), a \in \mathbb{F}_2
            \end{equation}
            for $j=0,1,\cdots,n-1$ and $i=1,2,\cdots,m$.
      \end{enumerate}
\end{itemize}
\end{algorithm}

\subsection{Stopping Criteria}\label{subsec:StoppingCriteria}
\subsubsection{Entropy-Based Stopping Criterion}\label{subsubsec:EntropyBased}
As the error-detection ability of short codes is usually weak, the entropy-based stopping criterion~\cite{Ma04} is used. The entropy-based stopping criterion is described as follows.

{\em The entropy-based stopping criterion for Algorithm~\ref{alg:TotallyDecoding}}: Before the iteration, we set a threshold $\epsilon > 0$ and initialize the entropy rate $h_{0}\left(\mathbfit{Y}\right)=0$, where $\mathbfit{Y}=\left( \mathbfit{Y}^{(0)}, \mathbfit{Y}^{(1)}, \cdots, \mathbfit{Y}^{(L+m-1)} \right)$ is the random vector corresponding to $\mathbfit{y}=\left( \mathbfit{y}^{(0)}, \mathbfit{y}^{(1)}, \cdots, \mathbfit{y}^{(L+m-1)} \right)$. For each iteration $I$, estimate the entropy rate of $\mathbfit{Y}$ by
\begin{equation}\label{eq:EntropyStoppingCriterion}
h_{I}\left(\mathbfit{Y}\right) = -\frac{1}{n(L+m)}\sum_{t=0}^{L+m-1}\sum_{j=0}^{n-1}\log\left(  P_{Y_j^{(t)}}^{(+\rightarrow|)}\left( y_j^{(t)} \right) \right),
\end{equation}
where,
\begin{eqnarray}\label{eq:Prob_y}
P_{Y_j^{(t)}}^{(+\rightarrow|)}\left( y_j^{(t)} \right) = \sum\limits_{a \in \mathbb{F}_2}
P_{C^{(t)}_{j}}^{(+\rightarrow|)}\left(a\right)
\cdot \mbox{Pr}\lbrace y_j^{(t)} | c_j^{(t)}=a\rbrace
\end{eqnarray}
and $P_{\mathbfit{C}^{(t)}}^{(+\rightarrow |)}\left( \mathbfit{c}^{(t)} \right)$ are computed at the node \fbox{+}. If $\left| h_{I}\left(\mathbfit{Y}\right) - h_{I-1}\left(\mathbfit{Y}\right) \right| \leq \epsilon$, exit the iteration.

{\em The entropy-based stopping criterion for Algorithm~\ref{alg:decoding}}: Before the iteration, we set a threshold $\epsilon > 0$ and initialize the entropy rate $h_{0}\left(\mathbfit{Y}^{(t)}\right)=0$, where $\mathbfit{Y}^{(t)}$ is the random vector corresponding to $\mathbfit{y}^{(t)}$. For each iteration $I$, estimate the entropy rate of $\mathbfit{Y}^{(t)}$ by
\begin{equation}\label{eq:EntropyStoppingCriterion}
h_{I}\left(\mathbfit{Y}^{(t)}\right) = -\frac{1}{n}\sum_{j=0}^{n-1}\log\left(  P_{Y_j^{(t)}}^{(+\rightarrow|)}\left( y_j^{(t)} \right) \right),
\end{equation}
where, $P_{Y_j^{(t)}}^{(+\rightarrow|)}\left( y_j^{(t)} \right)$ is computed by~(\ref{eq:Prob_y}) and $P_{\mathbfit{C}^{(t)}}^{(+\rightarrow |)}\left( \mathbfit{c}^{(t)} \right)$ are computed at the node \fbox{+}. If $\left| h_{I}\left(\mathbfit{Y}^{(t)}\right) - h_{I-1}\left(\mathbfit{Y}^{(t)}\right) \right| \leq \epsilon$, exit the iteration.

\subsubsection{Parity-Check-Based Stopping Criterion}\label{subsubsec:CRCBased}
To avoid the extra computational complexity caused by estimating the entropy rate, we may take a concatenated code as the basic code, where the outer code is a powerful error-detection code~(say cyclic redundancy check~(CRC) code) and the inner code is a short code. In this situation, the SISO algorithm for the basic code is performed by ignoring the constraint specified by the outer code. In the process of the iterative decoding, once the decoding output of the inner code is a valid codeword of the outer code, report a decoding success and exit the iteration.

\subsection{List Decoding after Iteration}\label{subsec:ListDecoding}
The use of error-detection codes for early stopping incurs a rate loss, however, it can be used to obtain extra coding gain by list decoding~\cite{Elias57,Wozencraft58}. In the case when the decoding fails after $I_{\max}$ iterations, the list decoding algorithm for the inner code takes $P_{\mathbfit{V}^{(t)}}^{(=\rightarrow C)}\left( \mathbfit{v}^{(t)} \right)$  as input and generates a list of outputs. Once one output in the list is found to be a valid codeword of the outer code, report a decoding success and exit the iteration.

\section{Performance Analysis}\label{sec:PerformanceAnalysis}
The objective of this section is to analyze the ``extra" coding gain over the basic code by the BMST system. Before doing this, we need to point out that the ``extra" coding gain may be negative in the high bit-error-rate~(BER) region due to the possible error propagation. Let $p_b = f_o(\gamma_b)$ be the performance function of the basic code $\mathscr{C}$, where $p_b$ is the BER and $\gamma_b \stackrel{\Delta}{=} E_b/N_0$ in dB. Since $\mathscr{C}$ is short, we assume that $p_b = f_o(\gamma_b)$ is available. For example, if $\mathscr{C}$ is a terminated convolutional code, the performance function under the {\em maximum a posteriori probability}~(MAP) decoding can be evaluated by performing the Bahl-Cocke-Jelinek-Raviv~(BCJR) algorithm~\cite{BCJR74}. Let $p_b = f_{\mbox{\small BMST}}(\gamma_b)$ be the performance function corresponding to the BMST system.

\subsection{Genie-Aided Lower Bound on BER}\label{subsec:UpperBound}
Let $\mathbfit{u} = (\mathbfit{u}^{(0)}, \mathbfit{u}^{(1)}, \cdots, \mathbfit{u}^{(L-1)})$ be the transmitted data. To derive the lower bound, we assume the MAP decoder for the BMST system, which in principle computes~(by Bayes' rule)
\begin{equation}\label{eq:lowerbound_deduction}
    {\rm Pr}\{u_j^{(t)} | \mathbfit{y}\} = \sum_{\mathbfit{ \tilde u}'} {\rm Pr}\{\mathbfit{\tilde u}' | \mathbfit{y}\} {\rm Pr}\{u_j^{(t)} | \mathbfit{\tilde u}', \mathbfit{y}\}
\end{equation}
for all $t$ and $j$, where the summation is over all $\mathbfit{\tilde u}' = (\mathbfit{\tilde u}^{(0)},  \cdots, \mathbfit{\tilde u}^{(t-1)}, \mathbfit{\tilde u}^{(t+1)}, \cdots, \mathbfit{\tilde u}^{(L-1)})$. We know that if ${\rm Pr}\{u_j^{(t)} | \mathbfit{y}\} > 0.5$, the decoding output is correct for this considered bit. In the meanwhile, we assume a \emph{genie-aided decoder}, which computes ${\rm Pr}\{u_j^{(t)} | \mathbfit{u}', \mathbfit{y}\}$ for all $t$
and $j$ with the transmitted data $\mathbfit{u}' = (\mathbfit{u}^{(0)}, \cdots, \mathbfit{u}^{(t-1)}, \mathbfit{u}^{(t+1)}, \cdots, \mathbfit{u}^{(L-1)})$ available.  Likewise, if ${\rm Pr}\{u_j^{(t)} | \mathbfit{u}', \mathbfit{y}\} > 0.5$, the decoding output is correct for this considered bit. For a specific $u_j^{(t)}$ and $\mathbfit{y}$, it is possible that ${\rm Pr}\{u_j^{(t)} | \mathbfit{u}', \mathbfit{y}\} < {\rm Pr}\{u_j^{(t)} | \mathbfit{y}\}$. However, the expectation
\begin{eqnarray}\label{eq:MutualINfo}
\mathbb{E}\left[ \log \frac{{\rm Pr}\{u_j^{(t)} | \mathbfit{u}',
\mathbfit{y}\}}{{\rm Pr}\{u_j^{(t)} | \mathbfit{y}\}} \right] = I
\left(U_j^{(t)}; \mathbfit{U}'  | \mathbfit{Y}\right)
\geq 0,
\end{eqnarray}
where $I \left(U_j^{(t)}; \mathbfit{U}'  | \mathbfit{Y}\right)$ is the conditional mutual information, implying that the genie-aided decoder performs statistically better than the MAP decoder of the BMST system. As a result, the BER performance can be lower-bounded by, taking into account the rate loss,
\begin{equation}\label{lowerbound}
    f_{{\small \rm BMST}}(\gamma_b) \geq f_{{\small \rm
    Genie}}(\gamma_b) = f_o(\gamma_b + 10\log_{10}(m+1) -
    10\log_{10}(1+m/L)),
\end{equation}
where the last equality holds from the fact that the data block $\mathbfit{u}^{(t)}$ is encoded and transmitted $m+1$ times from the perspective of the genie-aided decoder.

Furthermore, noticing that ${\rm Pr}\{\mathbfit{u}' | \mathbfit{y}\} \approx 1$ for the transmitted data block $\mathbfit{u}'$ in the low error rate region, we have from~(\ref{eq:lowerbound_deduction}) that ${\rm Pr}\{u_j^{(t)} |
\mathbfit{y}\} \approx {\rm Pr}\{u_j^{(t)} | \mathbfit{u}', \mathbfit{y}\}$ and hence can expect that \begin{equation}\label{eq:LowerBoundEquality}
    f_{\mbox{\small BMST}}(\gamma_b) \approx f_o(\gamma_b + 10\log_{10}(m+1) - 10\log_{10}(1+m/L))
\end{equation}
as $\gamma_b$ increases. That is, the maximum coding gain can be $10\log_{10}(m+1)$ dB for large $L$ in the low error rate region.

\subsection{Upper Bound on BER}\label{sec:LowerBound}
To upper-bound the BER performance, we present a method to compute the IOWEF of the BMST system. Let the IOWEF of the basic code $\mathscr{C}$ be given as
\begin{equation}
\label{eq:IOWEF_C}
B \left( X, Y \right) \triangleq \sum_{i,j} B_{i,j} X^{i} Y^{j},
\end{equation}
where $X$, $Y$ are two dummy variables and $B_{i,j}$ denotes the number of codewords having a Hamming weight $j$ when the corresponding input information sequence having a Hamming weight $i$. Similarly, denote by $A(X, Y)$ the IOWEF of the BMST system. We have
\begin{eqnarray}
A(X, Y) &=& \sum_{i,j} A_{i,j} X^{i} Y^{j} \nonumber \\
        &=& \sum_{\mathbfit{u}}X^{W_H(\mathbfit{u})}Y^{W_H(\mathbfit{c})} \nonumber \\
        &=& \sum_{\mathbfit{u}}\prod_{t = 0}^{L+m-1}X^{W_H(\mathbfit{u}^{(t)})}Y^{W_H(\mathbfit{c}^{(t)})},
\end{eqnarray}
where $W_H(\cdot)$ represents the Hamming weight and the summation is over all possible data sequences $\mathbfit{u}$ with $\mathbfit{u}^{(t)} = \mathbfit{0}$ for $t \geq L$. Since it is a sum of products, $A(X, Y)$ can be computed in principle by a trellis-based algorithm over the polynomial ring. For specific interleavers, the trellis has a state space of size $2^{mk}$. To make the computation tractable, we turn to an ensemble of BMST system by assuming that the interleavers are chosen independently and uniformly at random for each transmission block $\mathbfit{c}^{(t)}$. With this assumption, we can see that $W_H(\mathbfit{c}^{(t)})$ is a random variable that depends on the Hamming weights $\{W_H(\mathbfit{v}^{(i)}), t-m \leq i \leq t\}$.

In the following, we take $m = 1$ as an example to describe the algorithm for computing the IOWEF of the defined ensemble of the BMST system. We can see that $W_H(\mathbfit{c}^{(t)})$ is a random variable which is sensitive to neither $\mathbfit{v}^{(t-1)}$ nor $\mathbfit{v}^{(t)}$ but depends {\em only} on their Hamming weights $p = W_H(\mathbfit{v}^{(t-1)})$ and $q = W_H(\mathbfit{v}^{(t)})$. To be precise, we have
\begin{equation}
W_H(\mathbfit{c}^{(t)})=p+q-2r
\end{equation}
with probability
\begin{equation}\label{overlap_pro}
{\rm Pr} \{ W_H(\mathbfit{c}^{(t)})=p+q-2r \} =
\frac{ \binom{p}{r}\binom{n-p}{q-r} }{ \binom{n}{q} },
\end{equation}
where
\begin{equation}\label{overlap}
    r {=}
    \left\{
      \begin{array}{ll}
        0,1,\cdots,\min(p,q), & p+q \leq n \\
        p+q-n,\cdots,\min(p,q) , & p+q > n
      \end{array}
    \right..
\end{equation}

The trellis is time-invariant. At stage $t$, the trellis has $n+1$ states, each of which records the Hamming weight $W_H(\mathbfit{v}^{(t-1)})$. Emitting from each state there are $n+1$ branches, each of which corresponds to the Hamming weight $W_H(\mathbfit{v}^{(t)})$. To each branch $p\rightarrow q$, we assign a ``metric"
\begin{equation}
\gamma_{p\rightarrow q}=\sum_{r}{\rm
Pr}\{W_H(\mathbfit{c}^{(t)}) = p+q-2r\}\sum_{j}B_{j,q}X^{j}Y^{p+q-2r}.
\end{equation}
Then $A(X,Y)$ can be calculated recursively by performing a forward
trellis-based algorithm~\cite{Ma03} over the polynomial ring as follows.\vspace{0.15cm}
\begin{algorithm}{Computing IOWEF of BMST with $m=1$}\label{alg:IOWEF}
\begin{enumerate}
  \item Initialize $\alpha_0(p)=\sum_{j}B_{j,p}X^jY^p$,
        $p\in \{0,1,\cdots,n\}$.
  \item For $t = 0$, $1$, $\cdots$, $L-1$,
        \begin{equation}
          \alpha_{t+1}(q)=\sum_{p:p\rightarrow q}\alpha_t(p)\gamma_{p\rightarrow q},
          \nonumber
        \end{equation}
        where $q \in \{0,1,\cdots,n\}$.
  \item At time $L$, we have $A(X,Y)=\alpha_{L}(0)$.
\end{enumerate}
\end{algorithm}\vspace{0.15cm}

Given $A(X,Y)$, the upper bound for the BER of the BMST system can be calculated by an improved union bound~\cite{Ma13a}.

\textbf{Remark.}~For $m>1$, the computation becomes more complicated due to the huge number of trellis states $(n+1)^m$. Fortunately, as shown in~\cite{Ma13a}, truncated IOWEF suffices to give a valid upper bound, a fact that can be used to simplify the computation by removing certain states from the trellis.

\section{Numerical Results}\label{sec:Result}
In this section, we present BMST examples with different types of basic codes. All simulations are conducted by assuming BPSK modulation and AWGN channels. In all the examples, we set $I_{\max}=18$ as the maximum number of iterations. In the examples where the entropy-based stopping criterion is used, we set $\epsilon = 10^{-5}$ as the threshold. Without specification, the iterative sliding-window algorithm~(Algorithm~\ref{alg:decoding}) is used for decoding and S-random interleavers~\cite{Dolinar95}~(randomly generated but fixed) with parameter $S=\lfloor\sqrt{(n/4)}\rfloor$ are used for encoding. Here $\lfloor x \rfloor$ stands for the maximum integer that is not greater than $x$.

\subsection{Short Convolutional Codes as Basic Codes}\label{subsec:ConvolutionalCodes}
In this subsection, the BCJR algorithm is performed as the SISO decoding algorithm for basic codes and the entropy-based stopping criterion is used.

\begin{figure}[t]
    \centering
    \includegraphics[width=\figwidth]{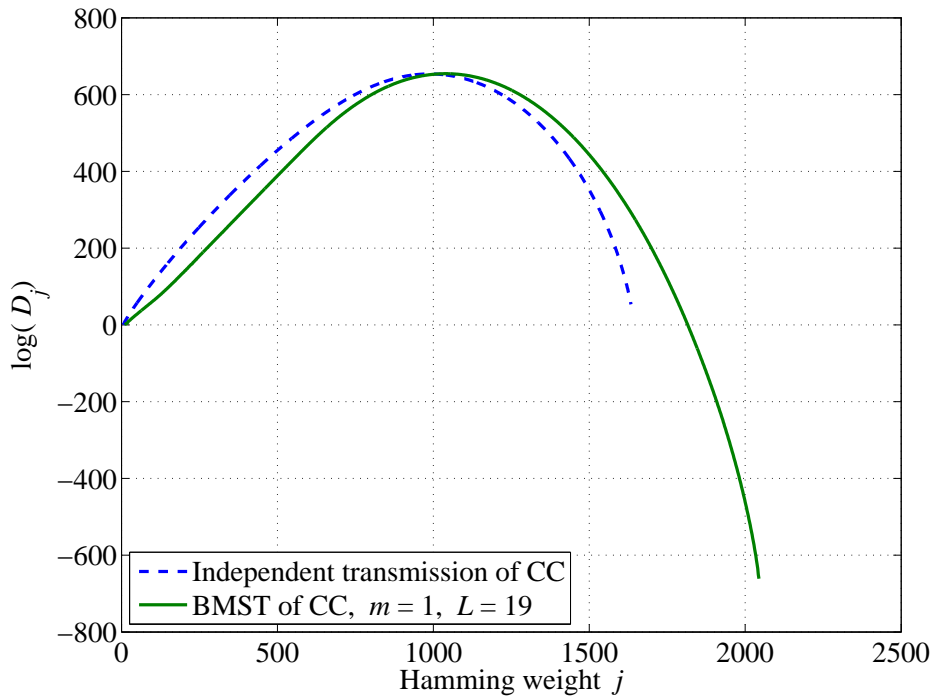}
    \caption{Comparison of the spectrum $\{D_j\}$ between the independent transmission system and the ensemble of the BMST system in Example~\ref{ex:MLDanalysis}. The basic code is a terminated systematic encoded 4-state $(2,1,2)$ convolutional code defined by the polynomial generator matrix $G(D)= [1, (1+D+D^2)/(1+D^2)]$. The BMST system encodes $L=19$ sub-blocks of data with memory $m = 1$.}
    \label{fig:WeightDistributionAnalysis}
\end{figure}
\begin{figure}[t]
    \centering
    \includegraphics[width=\figwidth]{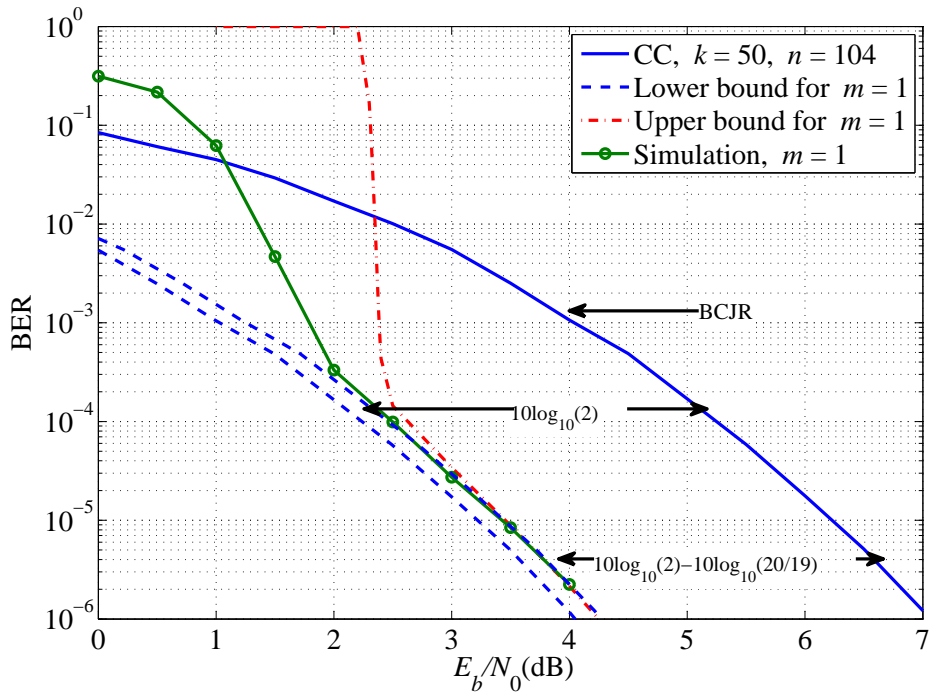}
    \caption{Performance of the BMST system in Example~\ref{ex:MLDanalysis}. The basic code is a terminated systematic encoded 4-state $(2,1,2)$ convolutional code defined by the polynomial generator matrix $G(D)= [1, (1+D+D^2)/(1+D^2)]$. The system encodes $L=19$ sub-blocks of data with memory $m = 1$. The decoding algorithm is performed after all 20 transmitted sub-blocks are received~(Algorithm~\ref{alg:TotallyDecoding}).}
    \label{fig:result_performance_analysis}
\end{figure}
\begin{example}\normalfont\label{ex:MLDanalysis}
The basic code $\mathscr{C}$ is a terminated systematic encoded 4-state $(2,1,2)$ convolutional code~(CC) defined by the polynomial generator matrix $G(D)= [1, (1+D+D^2)/(1+D^2)]$ with dimension $k = 50$ and length $n = 104$. We take $m =1, L=19$ for encoding. The decoding is performed after all $\mathbfit{y}^{(t)}$ are received~(Algorithm~\ref{alg:TotallyDecoding}). Fig.~\ref{fig:WeightDistributionAnalysis} shows the spectrum $\{D_j\}$ of the ensemble of the BMST system, where
\begin{equation}\label{eq:BitMultiplicity}
  D_{j} = \sum_{i=1}^{Lk}\frac{i}{Lk}A_{i,j}.
\end{equation}
For comparison, the spectrum of the independent transmission system~(with code book \\ $\{\left(\mathbfit{v}^{(0)}, \mathbfit{v}^{(1)}, \cdots, \mathbfit{v}^{(L-1)}, \mathbf{0} \right)\}$ instead of the BMST code book $\{\left( \mathbfit{c}^{(0)}, \mathbfit{c}^{(1)}, \cdots , \mathbfit{c}^{(L-1)}, \mathbfit{c}^{(L)}\right)\}$) is also shown in Fig.~\ref{fig:WeightDistributionAnalysis}. We can see that the spectrum of the BMST system has less number of codewords with small Hamming weights, indicating that the BMST system has potentially better performance than the independent transmission system. Simulation results are shown in Fig.~\ref{fig:result_performance_analysis}, which match well with the bounds in the high signal-to-noise ratio~(SNR) region. This also indicates that the iterative forward-backward algorithm is near optimal in the high SNR region.
\end{example}

\begin{figure}[t]
    \centering
    \includegraphics[width=\figwidth]{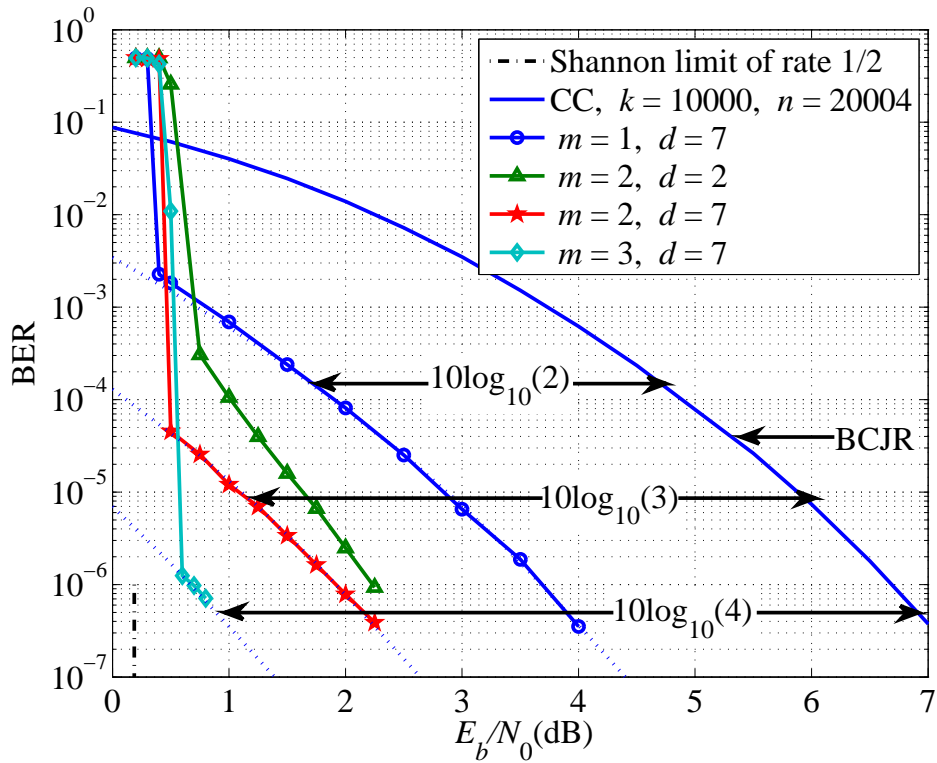}
    \caption{Performance of the BMST system in Example~\ref{ex:cc212}. The basic code is a terminated 4-state $(2,1,2)$ convolutional code defined by the polynomial generator matrix $G(D)=[1+D^2,1+D+D^2]$. The system encodes $L=1000$ sub-blocks of data and the iterative sliding-window decoding algorithm is performed, where the encoding memories and the decoding delays are specified in the legends.}
    \label{fig:result_convolutional}
\end{figure}
\begin{example}\label{ex:cc212}
The basic code $\mathscr{C}$ is a terminated 4-state $(2,1,2)$ convolutional code defined by the polynomial generator matrix $G(D)=[1+D^2,1+D+D^2]$ with $k=10000$ and $n=20004$. Simulations results for $L=1000$ are shown in Fig.~\ref{fig:result_convolutional}. We can see that 1)~given the encoding memory $m$, the performance can be improved by increasing the decoding delay $d$ and 2)~the performance in the high SNR region can be improved by increasing $m$.
\end{example}

\subsection{Short Block Codes as Basic Codes}\label{subsec:BlockCodes}
In this subsection, the BCJR algorithm is performed as the SISO decoding algorithm for basic codes and the entropy-based stopping criterion is used.

\begin{figure}[t]
    \centering
    \includegraphics[width=\figwidth]{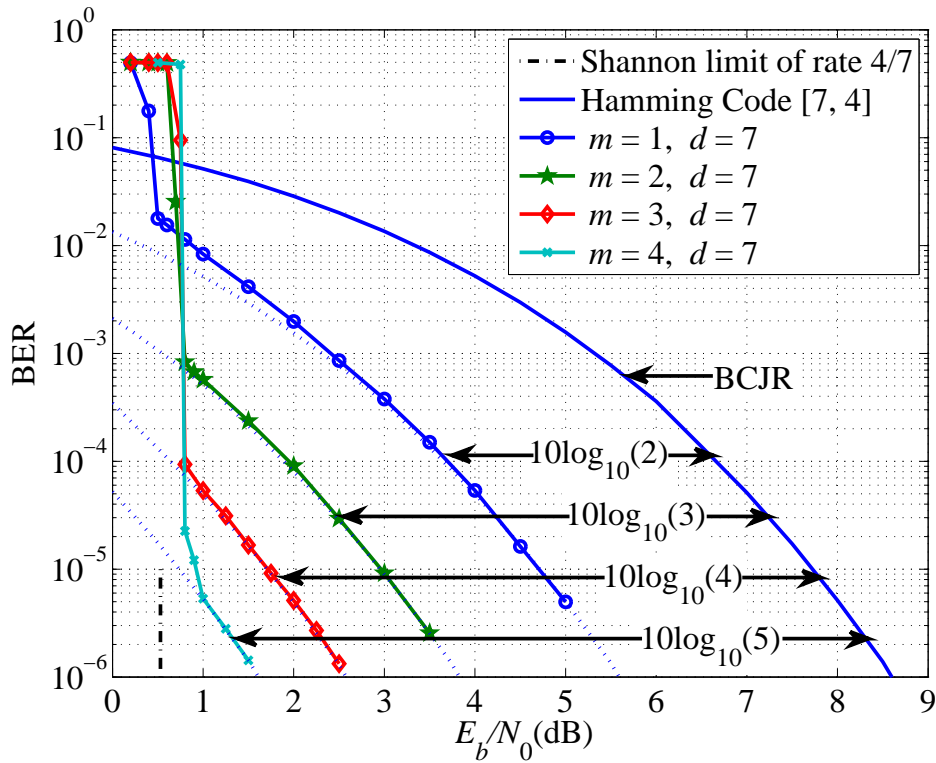}
    \caption{Performance of the BMST system in Example~\ref{ex:Hamming}. The basic code is the Cartesian product of Hamming code $[7,4]^{2500}$. The system encodes $L=1000$ sub-blocks of data and the iterative sliding-window decoding algorithm is performed, where the encoding memories and the decoding delays are specified in the legends.}
    \label{fig:result_Hamming}
\end{figure}
\begin{example}\label{ex:Hamming}
The basic code $\mathscr{C}$ is the Cartesian product of Hamming code $[7,4]^{2500}$ with $k=10000$ and $n=17500$. Simulation results for $L=1000$ and $d=7$ are shown in Fig.~\ref{fig:result_Hamming}. We can see that the BMST system of the Hamming code has a similar behavior to the BMST system of the convolutional code in Example~\ref{ex:cc212}. With $m=4$ and $d=7$, an extra coding gain of $6.7$~dB is obtained at BER $10^{-5}$.
\end{example}

\begin{figure}[t]
    \centering
    \includegraphics[width=\figwidth]{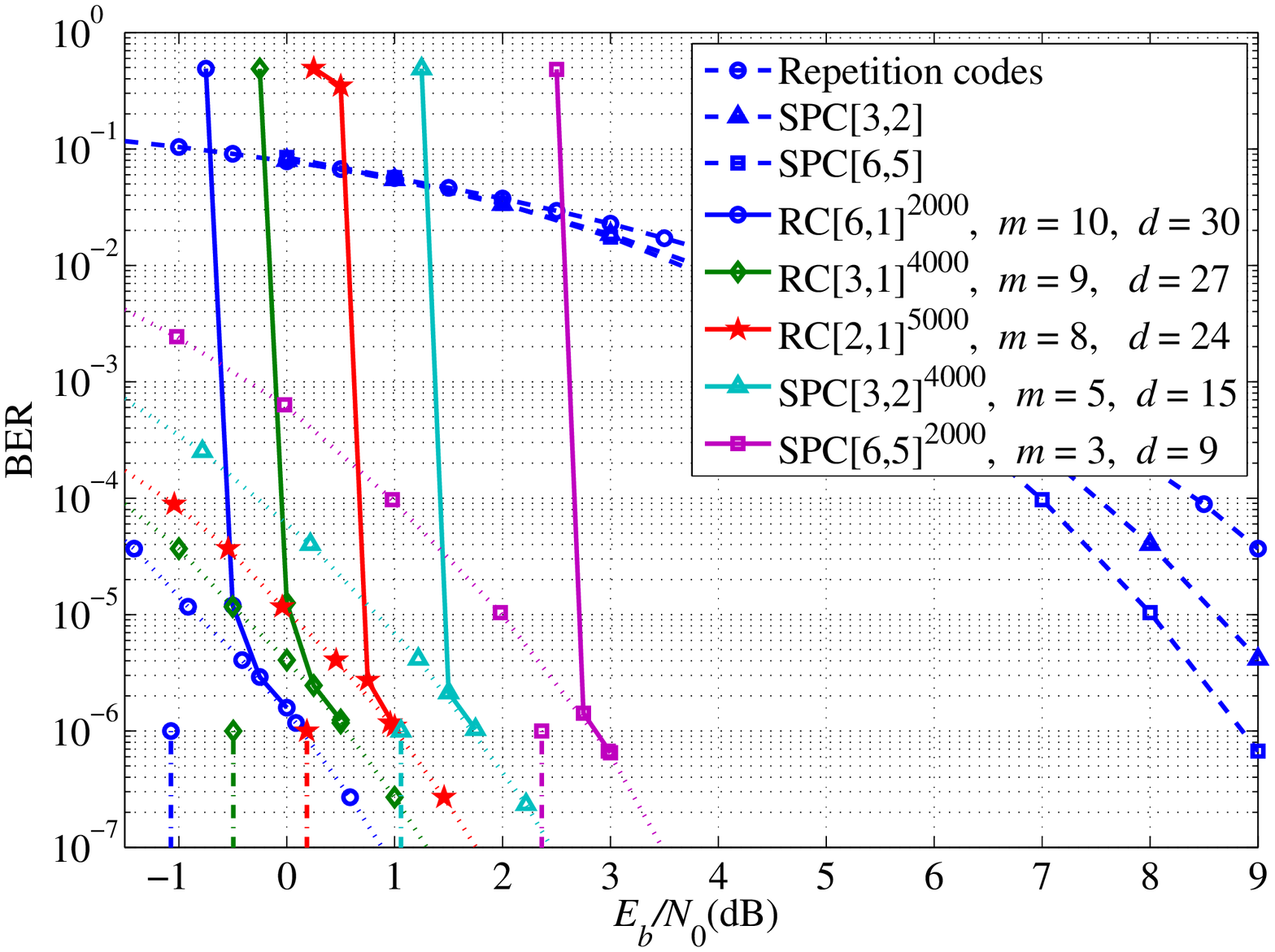}
    \caption{Performance of the BMST systems in Examples~\ref{ex:RepeatAndSingleParityCheckCode}. The basic code is either the Cartesian product of a repetition code or the Cartesian product of a single parity-check code. All systems encode $L=1000$ sub-blocks of data and the iterative sliding-window decoding algorithm is performed, where the encoding memories and the decoding delays are specified in the legends. The vertical dashed lines correspond to the respective Shannon limits.}
    \label{fig:resultRepeatAndSingleParityCheckCode}
\end{figure}
\begin{example}\label{ex:RepeatAndSingleParityCheckCode}
The basic code is either the Cartesian product of a repetition code~(RC), denoted by ${\rm RC}[n,1]^{N}$, or the Cartesian product of a single parity-check~(SPC) code, denoted by ${\rm SPC}[n,n-1]^{N}$. {Simulation results with $L=1000$ for all BMST systems are shown in Fig.~\ref{fig:resultRepeatAndSingleParityCheckCode}.} Also shown in Fig.~\ref{fig:resultRepeatAndSingleParityCheckCode} are the Shannon limits. More precisely, the Shannon limit of a code rate is depicted as a vertical dashed line, which shares the same mark with the solid performance curve of the given code rate. We can see that, given a short code, the corresponding Shannon limit can be approached using the BMST system by choosing properly the encoding memory and the decoding delay. There is about 0.5 dB away from the respective Shannon limit at BER=$10^{-5}$ for all BMST systems given in Fig.~\ref{fig:resultRepeatAndSingleParityCheckCode}.
\end{example}

\subsection{Concatenation of CRC Codes and Short Convolutional Codes as Basic Codes}\label{subsec:BlockCodes}
If a concatenated code with a powerful error-detection outer code is used as the basic code, we can use the parity-check-based stopping criterion for early stopping. In this case, the SISO algorithm for the basic code is performed by ignoring the outer code. To improve the performance, a list decoding can be implemented after the iteration.

\begin{figure}[t]
\centering
\includegraphics[width=\figwidth]{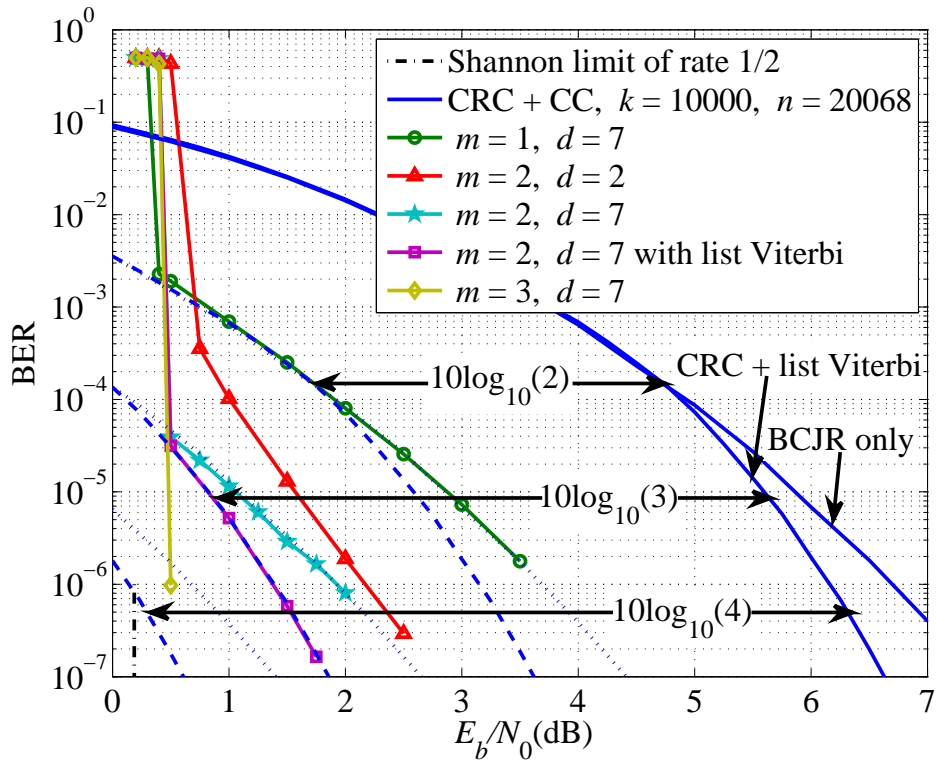}
\caption{Performance of the BMST system in Example~\ref{ex:crccc212}. The basic code is a concatenated code, where the outer code is a 32-bit CRC code and the inner code is a terminated 4-state $(2,1,2)$ convolutional code defined by the polynomial generator matrix $G(D)=[1+D^2,1+D+D^2]$. The system encodes $L=1000$ sub-blocks of data and the iterative sliding-window decoding algorithm is performed, where the encoding memories and the decoding delays are specified in the legends.}
\label{fig:crccc212}
\end{figure}
\begin{example}\label{ex:crccc212}
The basic code $\mathscr{C}$ is a concatenated code with $k=10000$ and $n=20068$, where the outer code is a 32-bit CRC code and the inner code is a terminated 4-state $(2,1,2)$ convolutional code defined by the polynomial generator matrix $G(D)=[1+D^2,1+D+D^2]$. Simulation results for $L=1000$ are shown in Fig.~\ref{fig:crccc212}. We can see that the simulation results are similar to those in Example~\ref{ex:cc212}.  The BER curves match well with the lower bounds derived from the BCJR-only curves but diverge from the bounds derived from the list-Viterbi~(with list size 2) curves. The reason is as follows. During the iterative sliding-window decoding of BMST systems, the CRC code serves only as an error-detection code to exit the iteration at the right time and is less useful to enhance the error performance. To verify this, a list decoding with list size $2$ is implemented after the failure of the iterative sliding-window decoding. As expected, the simulation results match well with the bounds derived from the list-Viterbi curves. For an example, see the curve in Fig.~\ref{fig:crccc212} with $m = 2$ and $d = 7$.
\end{example}

\section{On the Universality of the BMST}\label{sec:Universality}
\subsection{Sketch of the Performance Curve}\label{subsec:generalcurve}
\begin{figure}[t]
    \centering
    \includegraphics[width=\figwidth]{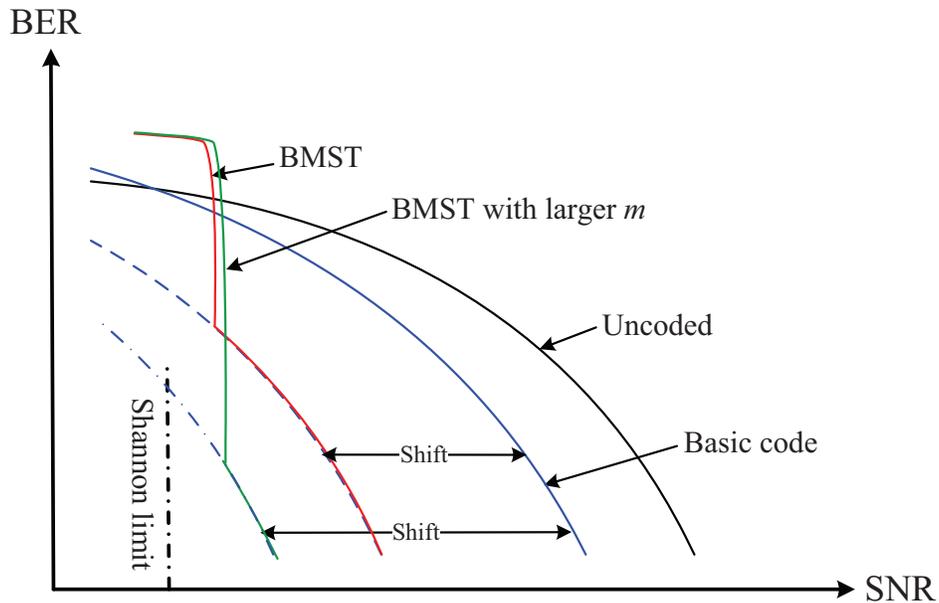}
    \caption{Sketches of the performance curves of a general BMST system.}
    \label{fig:GeneralBERcurve}
\end{figure}
We have conducted lots of simulations for BMST systems with a variety of basic codes while only some of them are presented in this paper due to the space limit. We have found that all simulations deliver performance curves that have similar behavior. That is, the performance curve drops down to the derived genie-aided lower bound as the decoding delay increases. Let $\mathscr{C}[n, k, d_{\min}]$ be the basic code, which is either a terminated convolutional code with a short constraint length or a Cartesian product of a short block code. In either case, we assume that $n$ is large enough. Fig.~\ref{fig:GeneralBERcurve} sketches the performance curves for a general BMST system. Compared with the uncoded system, as the SNR increases, the basic code has an asymptotic coding gain~(ACG) upper-bounded by $10\log_{10}(kd_{\min}/n)$ dB. The performance curve of the BMST system with encoding memory $m$ is lower-bounded by shifting to left that of the basic code by $10\log_{10}(1+m)$. Hence we have an extra ACG of $10 \log_{10}(1+m)$ dB. As the encoding memory increases, the curve of the performance lower bound shifts to left further. However, the waterfall part of the simulated curve {\em may} shift to right a little bit due to the error propagation. Because of the same reason, the BMST system performs worse than the basic code in the low SNR region.

\subsection{Nonlinear Codes as Basic Codes}\label{subsec:nonlinearcode}
From both the encoding process and the decoding process of the BMST system, we can see that the linearity of the basic code plays no essential roles. What we need is an encoding algorithm as well as an SISO decoding algorithm for the basic code. Hence, from the theoretical point of view, we are interested in the performance of BMST system with a nonlinear basic code. The advantage of the use of the nonlinear code is that the same coding gain may be obtained over the uncoded system with a less encoding memory provided the nonlinear basic code is better than the comparable linear basic code~\cite{Macwilliams77}. The disadvantage is that the table look-up encoding algorithm and the brute-force SISO decoding algorithm may be required for a general nonlinear basic code.

In the following example, we show that BMST can also be applied to nonlinear codes. Following~\cite{Macwilliams77}, a nonlinear binary code $\left(n, M, d_{\min}\right)$ is defined as a set of $M$ binary vectors of length $n$, any two of which have a Hamming distance at least $d_{\min}$ and some two of which have a Hamming distance $d_{\min}$. The code rate is $\log_2(M)/n$ and the ACG is upper-bounded by $10\log_{10}(d_{\min}\log_2(M)/n)$ dB. If necessary, the word-error rate~(WER) is used to measure the performance. The input to the encoder is an index~(carrying information) for some codeword in the nonlinear basic code. The table look-up encoding algorithm for the basic code is implemented in Algorithm~\ref{alg:encoding}. The brute-force MAP decoding algorithm based on Bayes' rule is implemented as the SISO decoding algorithm for the nonlinear basic code in the iterative sliding-window decoding algorithm. The entropy-based stopping criterion is used.

\begin{figure}[t]
    \centering
    \includegraphics[width=\figwidth]{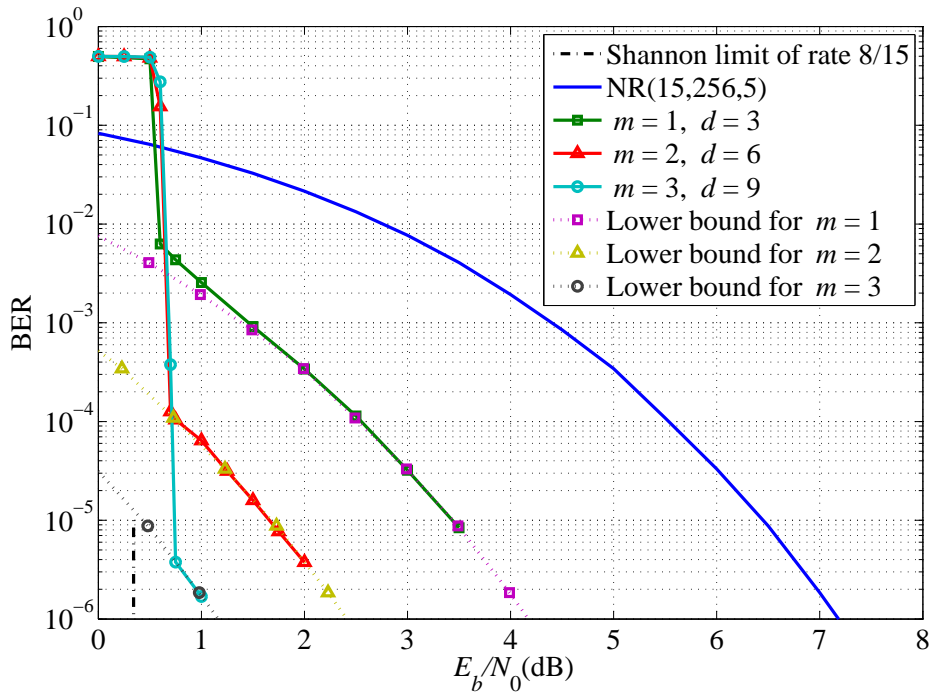}
    \caption{Performance of the BMST system in Example~\ref{ex:NonliearCode}. The basic code is the Cartesian product of the optimum Nordstrom-Robinson nonlinear code $(15, 256, 5)^{800}$. The system encodes $L=1000$ sub-blocks of data and the iterative sliding-window decoding algorithm is performed, where the encoding memories and the decoding delays are specified in the legends.}
    \label{fig:NonlinearCode}
\end{figure}
\begin{example}\label{ex:NonliearCode}
The basic code is the Cartesian product of the optimum Nordstrom-Robinson~(NR) nonlinear code $(15, 256, 5)^{800}$~\cite{Nordstrom67,Preparata68}. As pointed out in~\cite{Macwilliams77}, the Nordstrom-Robinson nonlinear code $(15, 256, 5)$ contains (at least) twice as many codewords as any linear code with the same length and minimum distance. {Simulation results for $L=1000$ are shown in Fig.~\ref{fig:NonlinearCode}.} We can see that the performance curves of the BMST with this nonlinear code match well with the corresponding lower bounds in the high SNR region, which are consistent with sketches as shown in Fig.~\ref{fig:GeneralBERcurve} for the general BMST system.
\end{example}

\subsection{Concatenated Codes as Basic Codes}\label{subsec:ConcatenatedCodes}
An inevitable but interesting question is whether the BMST is applicable to long codes. In this subsection, we present a BMST system with a concatenated code as the basic code, where the outer is a Reed-Solomon~(RS) and the inner code is a convolutional code. The issue of this system is that no efficient SISO decoding algorithm for the basic code exists~\footnote{Hence, we do not have the simple genie-aided lower bound in this case.}. As a trade-off, we implement the iterative sliding-window decoding algorithm by ignoring the existence of the outer code, which is used only for removing the residual errors and stopping the iterations at the right time. For each iteration, the outer decoder~(the Berlekamp-Massey~(BM)~\cite{Berlekamp68,Massey69} algorithm) is performed. Whenever it is  successful, the estimated data are output and the iteration is stopped.

\begin{figure}[t]
    \centering
    \includegraphics[width=\figwidth]{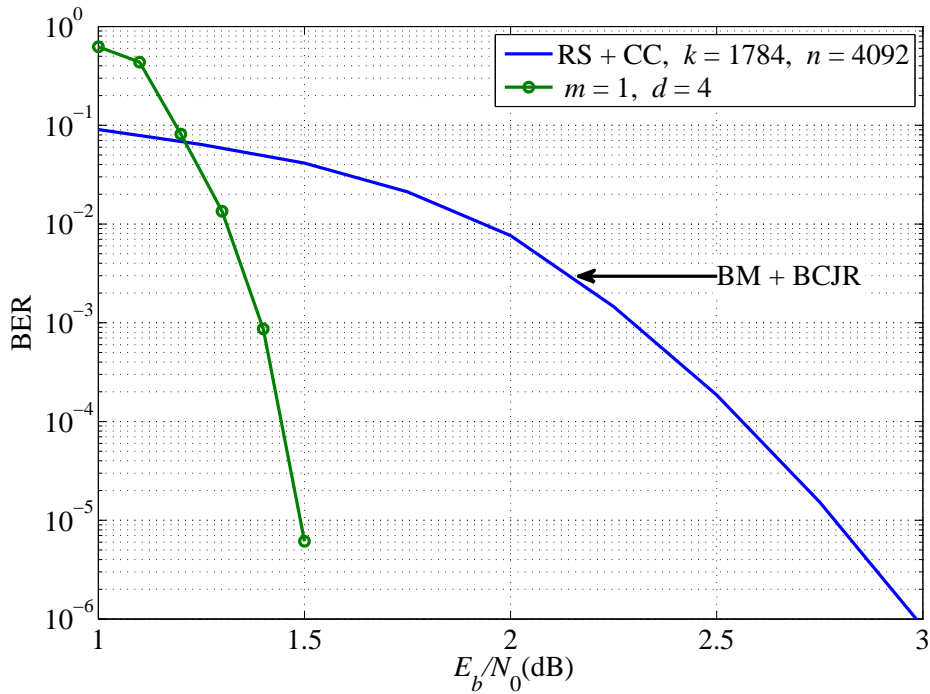}
    \caption{Performance of the BMST system in Example~\ref{ex:RSCC216}. The basic code is the CCSDS standard code, where the outer code is a [255, 223] RS code over $\mathbb{F}_{256}$ and the inner code is a terminated 64-state $(2,1,6)$ convolutional code defined by the polynomial generator matrix $G(D)=[1+D+D^2+D^3+D^6, 1+D^2+D^3+D^5+D^6]$. The system encodes $L=100$ sub-blocks of data with memory $m = 1$ and the iterative sliding-window decoding algorithm is performed with decoding delay $d=4$.}
    \label{fig:result_CCSDS_RSCC}
\end{figure}
\begin{example}\label{ex:RSCC216}
The basic code $\mathscr{C}$ is the Consultative Committee on Space Data System (CCSDS) standard code~\cite{CCSDS11} with $k=1784$ and $n=4092$, where the outer code is a [255, 223] RS code over $\mathbb{F}_{256}$ and the inner code is a terminated 64-state $(2,1,6)$ convolutional code defined by the polynomial generator matrix $G(D)=[1+D+D^2+D^3+D^6, 1+D^2+D^3+D^5+D^6]$. The RS code not only removes the possible residual errors after the iterative sliding-window decoding of the inner code but also ensures~(with high probability) the correctness of successfully decoded codewords.\footnote{The mis-correction probability can be analyzed in a similar way to that given in~\cite{Ma11x}.} The simulation results with $L=100$, $m=1$, and $d=4$ are shown in Fig.~\ref{fig:result_CCSDS_RSCC}. We can see that, although we are unable to simulate the performance in the extremely low BER region, the extra coding gain is about $1.3$~dB at BER $10^{-5}$. Also notice that the BMST system performs worse than the basic code in the high BER region due to the error propagation.
\end{example}

\section{Conclusion}\label{sec:Conclusion}
In this paper, we presented more details about the block Markov superposition transmission~(BMST), a construction of big convolutional codes from short codes. The encoding process can be as fast as the short code, while the decoding has a fixed delay. The coding gain of the BMST system is analyzed and verified by simulations. A nice property of the BMST is that its performance in the high SNR region can be approximately predicted. With several examples, we show that the BMST is a simple and general method for obtaining extra coding gain in the low BER region over short codes. With repetition codes and single parity-check codes as basic codes, the BMST system can approach the Shannon limit at BER $10^{-5}$ within 0.5 dB for a wide range of code rates.



\section*{Acknowledgment}
The authors are grateful to Prof. Baoming Bai from Xidian University for useful discussions.

\bibliographystyle{IEEEtran}


\balance

\end{document}